\documentclass[nonacm, anonymous=false, sigconf]{acmart}

\renewcommand\footnotetextcopyrightpermission[1]{} 
\newcommand{\argmax}{\mathop{\mathrm{argmax}}} 
\newcommand{\norm}[1]{\left\lVert #1 \right\rVert}

\usepackage{booktabs} 
\usepackage{pdfpages}
\usepackage{graphicx}
\usepackage{textcomp}
\usepackage{xcolor}
\usepackage{adjustbox}
\usepackage{multirow}
\usepackage{booktabs}
\usepackage{rotating}
\usepackage[inline]{enumitem}
\usepackage[para]{threeparttable}
\usepackage{etoolbox}
\usepackage{enumitem}
\usepackage{balance}
\usepackage{microtype}
\usepackage{amssymb,mathtools}
\usepackage{float}


\AtBeginDocument{
	\providecommand\BibTeX{{%
			\normalfont B\kern-0.5em{\scshape i\kern-0.25em b}\kern-0.8em\TeX}}}

\begin{document}
	
	\title[A Blast From the Past]{A Blast From the Past: Personalizing Predictions of Video-Induced Emotions using Personal Memories as Context}

	\author{Bernd Dudzik}
	\authornote{This is the corresponding author}
	\affiliation{%
		\institution{Delft University of Technology}
		\streetaddress{Van Mourik Broekmanweg 6}
		\city{Delft}
		\state{South Holland}
		\postcode{2628 XE}
		\country{The Netherlands}}
	\email{B.J.W.Dudzik@tudelft.nl}
	
	\author{Joost Broekens}
	\affiliation{%
		\institution{Delft University of Technology}
		\streetaddress{Van Mourik Broekmanweg 6}
		\city{Delft}
		\state{South Holland}
		\postcode{2628 XE}
		\country{The Netherlands}}
	\email{D.J.Broekens@tudelft.nl}
	
	\author{Mark Neerincx}
	\affiliation{%
		\institution{Delft University of Technology}
		\streetaddress{Van Mourik Broekmanweg 6}
		\city{Delft}
		\state{South Holland}
		\postcode{2628 XE}
		\country{The Netherlands}}
	\email{M.A.Neerincx@tudelft.nl}
	\author{Hayley Hung}
	\affiliation{%
		\institution{Delft University of Technology}
		\streetaddress{Van Mourik Broekmanweg 6}
		\city{Delft}
		\state{South Holland}
		\postcode{2628 XE}
		\country{The Netherlands}}
	\email{H.Hung@tudelft.nl}
	
	\renewcommand{\shortauthors}{Dudzik et al.}

	\begin{abstract}
		A key challenge in the accurate prediction of viewers' emotional responses to video stimuli in real-world applications is accounting for person- and situation-specific variation. An important contextual influence shaping individuals' subjective experience of a video is the personal memories that it triggers in them. Prior research has found that this memory influence explains more variation in video-induced emotions than other contextual variables commonly used for personalizing predictions, such as viewers' demographics or personality. In this article, we show \begin{enumerate*} \item that automatic analysis of text describing their video-triggered memories can account for variation in viewers' emotional responses, and \item that combining such an analysis with that of a video's audiovisual content enhances the accuracy of automatic predictions \end{enumerate*}. 
		We discuss the relevance of these findings for improving on state of the art approaches to automated affective video analysis in personalized contexts.  
		
	\end{abstract}
	
	\maketitle
	
	\section{Introduction}
	The experience of specific feelings and emotional qualities is an essential driver for people to engage with media content, e.g. for entertainment or to regulate their mood \cite{Bartsch2012}. For this reason, research on \textit{Video Affective Content Analysis (VACA)} attempts to automatically predict how people emotionally respond to videos \cite{Baveye2018}. This has the potential to enable media applications to present video content that reflects the emotional preferences of their users \cite{Hanjalic2005}, e.g. through facilitating emotion-based content retrieval and recommendation, or by identifying emotional highlights within clips. Existing VACA approaches typically base their predictions of emotional responses mostly on a video's audiovisual data \cite{Wang2015}.
	
	In this article, we argue that many VACA-driven applications can benefit from emotion predictions that also incorporate viewer- and situation-specific information as context for accurate estimations of individual viewers' emotional responses. Considering context is essential, because of the inherently subjective nature of human emotional experience, which is shaped by a person's unique background and current situation \cite{Greenaway2018}. As such, emotional responses to videos can be drastically different across viewers, and even the feelings they induce in the same person might change from one viewing to the next, depending on the viewing context \cite{Soleymani2014}. Consequently, to achieve affective recommendations that meaningfully match a viewer's current desires, e.g., to see an "amusing" video, these will likely need some sensitivity for the conditions under which any potential video will create this experience specifically for him/her. Therefore, addressing variation in emotional responses by personalizing predictions is a vital step for progressing VACA research (see the reviews by Wang et al. \cite{Wang2015} and Baveye et al. \cite{Baveye2018}). 
	
	Despite the potential benefits, existing research efforts still rarely explore incorporating situation- or person-specific context to personalize predictions. Reasons for this include that it is both difficult to \textit{conceptualize} context (identifying essential influences to exploit in automated predictions), as well as to \textit{operationalize} it (obtaining and incorporating relevant information in a technological system) \cite{Hammal2015}. Progress towards overcoming these challenges for context-sensitive emotion predictions requires systematic exploration in computational modeling activities, guided by research from the social sciences \cite{Dudzik2019}.
	
	Here, we contribute to these efforts by exploring the potential to account for personal memories as context in automatic predictions of video-induced emotions. Findings from psychology indicate that audiovisual media are potent triggers for personal memories in observers \cite{McDonald2015, Belfi2016}, and these can evoke strong emotions upon recollection \cite{Janata2007}. Once memories have been recollected, evidence shows that they possess a strong ability to elicit emotional responses, hence their frequent use in emotion induction procedures \cite{Mills2014}. Moreover, the affect associated with media-triggered memories -- i.e., how one feels about what is being remembered -- has a strong connection to its emotional impact \cite{Baumgartner1992a, Dudzik2020p1}. These findings indicate that by accessing the emotion associated with a triggered memory, we are likely to be able to obtain a close estimate of the emotion induced by the media stimulus itself. Moreover, they underline the potential that accounting for the emotional influence of personal memories holds for applications. They may enable technologies a more accurate overall reflection of individual viewers' emotional needs in affective video-retrieval tasks. However, they might also facilitate novel use-cases that target memory-related feelings in particular, such as recommending nostalgic media content \cite{Cosley2012}. 
	
	One possible way to address memory-influences in automated prediction is through the analysis of text or speech data in which individuals explicitly describe their memories. Prior research has revealed that people frequently disclose memories from their personal lives to others \cite{Rime2002, Walker2009}, likely because doing so is an essential element in social bonding and emotion regulation processes \cite{Bluck2005}. There is evidence that people share memories for similar reasons on social media \cite{Caci2019}, and that they readily describe memories triggered in them by social media content \cite{Cosley2012}. Moreover, predicting the affective state of authors of social media text \cite{Mohammad2018}, as well as text analysis of dialog content \cite{Chatterjee2019}, are active areas of research. However, apart from the work of Nazareth et al. \cite{Nazareth2019}, we are not aware of specific efforts to analyze memories. Together, these findings indicate that it may be feasible to both (1) automatically extract emotional meaning from free-text descriptions of memories and (2) that such descriptions may be readily available for analysis by mining everyday life speech or exchanges on social media. This second property may also make memory descriptions a useful source of information in situations where other data may be unavailable due to invasiveness of the required sensors -- e.g., when sensing viewers' facial expressions or physiological responses. Motivated by the potential of analyzing memories for supporting affective media applications, we present the following two contributions:
	\begin{enumerate}
		\itemsep 0em	
		\item we demonstrate that it is possible to explain variance in viewers' emotional video reactions by automatically analyzing free-text self-reports of their triggered memories, and
		\item we quantify the benefits for the accuracy of automatic predictions when combining both the analysis of videos' audiovisual content with that of viewers' self-reported memories
	\end{enumerate}

	\section{Background and Related Work} 
	\subsection{Video Affective Content Analysis} 
	
	With respect to the use of context, existing work can be categorized into two types: 
	\begin{enumerate*} 
		\item \textit{context-free VACA} and
		\item \textit{context-sensitive VACA}
	\end{enumerate*}.
	
\begin{figure}
		\includegraphics[width=\linewidth, trim={0em 0em 0em 0em},clip]{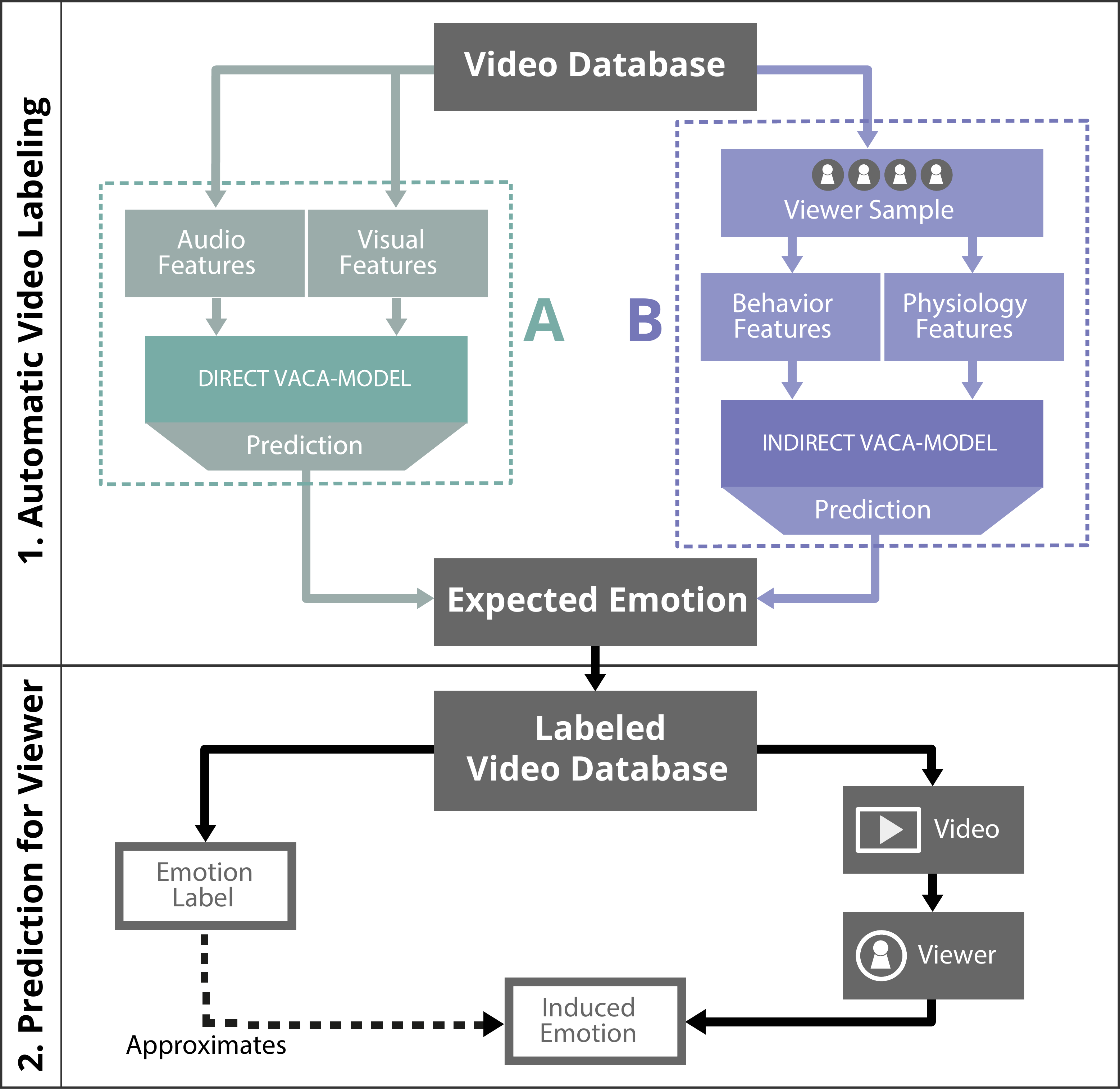}
		\vspace{-2em}
		\caption{Overview about the Context-Free VACA approach -- A: Direct VACA; B: Indirect VACA}
		\label{fig:sec2_VACA-CFree}
		\vspace{-1.5em}
	\end{figure}

	\paragraph{Context-Free VACA:} Works belonging to this type simplify the task of emotion prediction by making the working assumption that every video has quasi-objective emotional content \cite{Wang2015}. Traditionally, researchers define this content as the emotion that a stimulus results in for a \emph{majority} of viewers (i.e. its \textit{Expected Emotion} \cite{Baveye2018}). The goal of VACA technology then is to automatically provide a single label for a video representing its expected emotional content, while ignoring variation in emotional responses that a video might elicit. Existing technological approaches for this task primarily consist of machine learning models trained in a supervised fashion on human-annotated corpora \cite{Wang2015}. The ground truth for expected emotions is formed by aggregating the individual emotional responses from multiple viewers for the same video (e.g., by taking the mean or mode across the distribution of their responses). These models can then be used to automatically label entire databases of videos with tags representing their emotional content. The whole process of prediction of an individual viewer's emotional response using context-free VACA consists of two primary stages, as shown in the graphical overview displayed in \textit{Figure \ref{fig:sec2_VACA-CFree}}.  \begin{enumerate*} \item using a pre-trained VACA model to automatically label any video in a database of interest (with its expected emotion), and then \item relying on the video label to be a plausible approximation for the specific emotion that any individual viewer experiences (i.e., his/her \textit{Induced Emotion}) \end{enumerate*}. 
	
	There are two groups of technological approaches to automatically label videos in this way, differing in the information that they rely on as input for their predictions \cite{Wang2015}. The first, and traditionally most widespread approach is \emph{Direct VACA} (see region \textit{A} in \ref{fig:sec2_VACA-CFree}), which exclusively uses features derived from the audio and visual tracks of the video stimulus itself as the basis for predictions. The second is \emph{Indirect VACA}, which is looking into automated approaches for analyzing the spontaneous behavior displayed by viewers to label videos without having to ask them for a rating. In essence, this approach uses measures of physiological responses or human behavior (e.g. \cite{Koelstra2012, Soleymani2012-DB}) from a sample audience to predict the expected emotion of a population for a video (see region \textit{B} in \textit{Figure \ref{fig:sec2_VACA-CFree}}). As such, this endeavor is closely connected to the broader research on emotion recognition from human physiology and behavior in affective computing (see D'mello and Kory \cite{Dmello2015} for a recent overview). Unfortunately, during the writing of this article, we found that the conceptual differences between the two research efforts are not necessarily clearly defined. Therefore, in this paper, we define \emph{emotion recognition} to be approached using only measures of an individual's behavior or physiology to predict his/her specific emotional response to a video (e.g., as in Shukla et al. \cite{Shukla2017}). In contrast to this, Indirect VACA methods (also known as \textit{Implicit Tagging} \cite{Soleymani2012a}) collect behavioral or physiological data from multiple viewers in response to videos and use an aggregate of these measurements as input to label videos with their \emph{expected} emotion automatically. 
	
	Context-free approaches intentionally blend out issues of subjectivity and variation. In essence, they rely entirely on the \emph{expected} emotion-labels to be reasonable approximations for the emotional responses of an individual viewer to the video, independent of who he or she is, or in what situation they are watching the video. Because the prediction target is a video-specific aggregate, context-free affective content can only be valid if stimuli display only a small amount of within-video variation in responses. Creators of corpora for VACA modeling have typically enforced this by selecting only stimulus videos for which a low degree of induced emotional variation was already observed (e.g., \cite{Koelstra2012, Abadi2015, McDuff2017}). Nevertheless, in naturalistic viewing conditions, such strongly homogeneous responses are unrealistic \cite{Soleymani2014}. We argue that variation in emotional impact is the norm rather than the exception. Consequently, VACA research needs to expand its notion of affective content from being quasi-objective to inherently subjective and context-dependent if it wants to make predictions that meaningfully reflect individual's media experiences in the real world.

	\paragraph{Context-sensitive VACA:} We consider approaches as context-sensitive VACA when they 
	\begin{enumerate*}
		\item attempt to predict an individual viewer's affective responses to a video (i.e. the \textit{Induced Emotion} \cite{Baveye2018}), and
		\item rely on both the analysis of a video's audiovisual content and the context for this task.   
	\end{enumerate*} 
	
\begin{figure}
		\includegraphics[width=\linewidth, trim={0em 0em 0em 0em},clip]{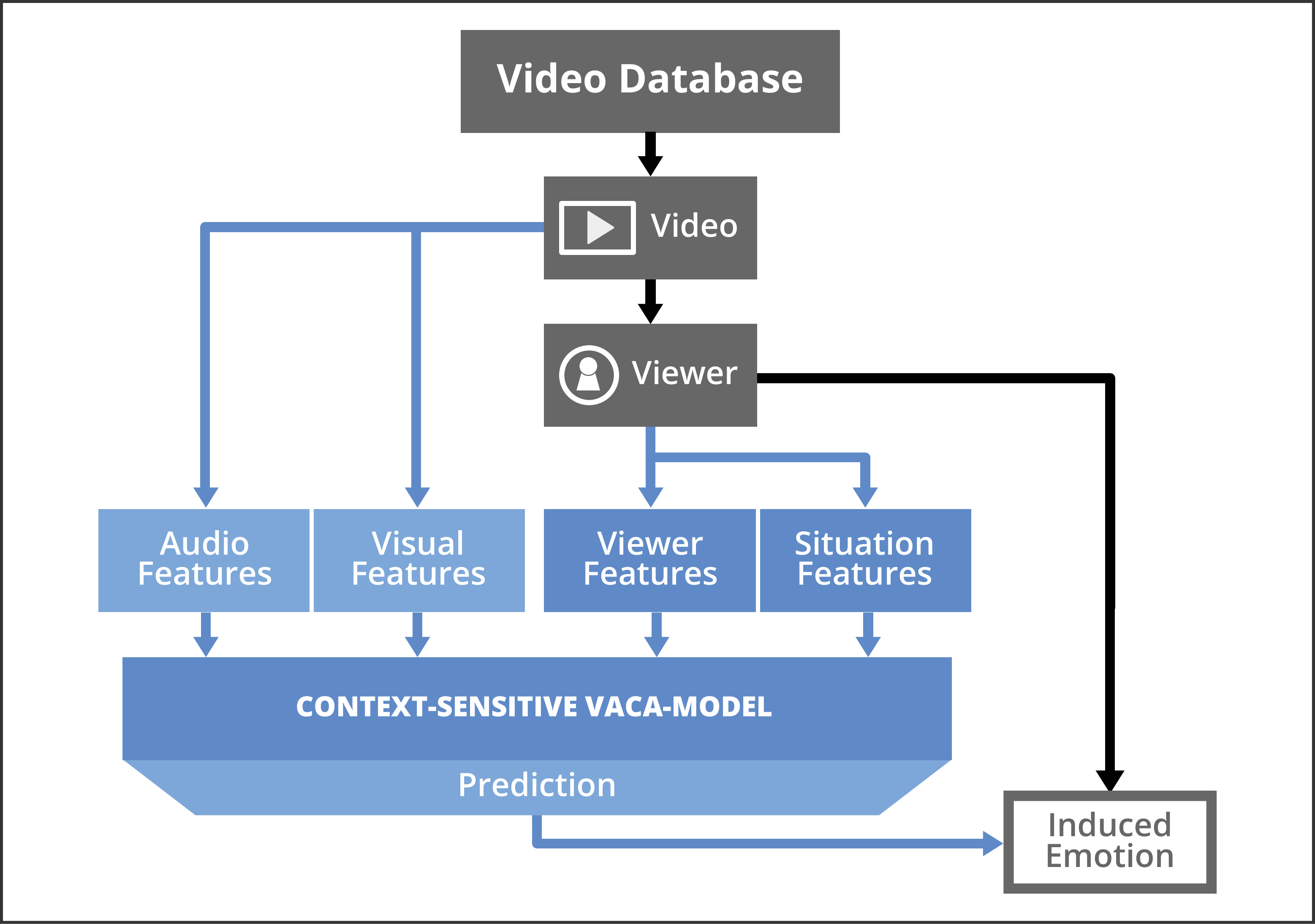}
		\vspace{-2em}
		\caption{Context-Sensitive VACA approaches}
		\label{fig:sec2_VACA-CSen}
		\vspace{-1em}
	\end{figure}

	Existing works can further be distinguished according to the type of context that they use:  
	
	\vspace{-0.5em}
	\begin{itemize}
		\itemsep 0em	
		\item \textit{Viewer-specific context} refers to properties and traits of individual viewers aimed at accounting for variation in video-induced emotions between different individuals (e.g. demographics, personality), while
		
		\item \textit{Situation-specific context} denotes information that is temporarily relevant for predicting a viewer's emotional responses. It covers influences that stretch over multiple videos viewed by the same person in succession (e.g. the type of social setting in which viewing takes place), or that may be specific to only a single instance of viewing (such as a triggered memory).
	\end{itemize} 
	\vspace{-0.5em}
	See \textit{Figure \ref{fig:sec2_VACA-CSen}} for a schematic overview of context-sensitive VACA.

	According to this coarse categorization, initial efforts have explored the impact of viewer-specific context, capturing their personality and cultural background \cite{Guntuku2015, Scott2016}. Other research has touched on situational properties, such as mood or the time of day at which someone watches a video \cite{Soleymani2014}. It has also explored the impact of viewing in a group compared to being alone \cite{Zhu2015}). Finally, measurements of an individual viewer's behavior or physiology while watching a video can be considered fine-grained information about their current situation and is explored extensively in existing computational approaches (e.g., \cite{Wang2014, Koelstra2012}). 
	
	\vspace{-1mm}
	\subsection{Representing Video-Induced Emotions}
	Researchers in psychology do not universally agree upon a single system to formally categorize the various states and experiences that make up human emotion. Instead, there exist competing schemes and taxonomies, many of which have been used to represent video-induced emotion \cite{Baveye2018}. These fall broadly into two distinct groups: categorical and dimensional schemes. Categorical schemes describe human emotion in terms of a set of discrete emotional states, such as happiness or anger. In contrast to this, dimensional frameworks describe subjective emotional experiences as points in a continuous, multidimensional space. Each dimension in the scheme intends to capture a vital characteristic to differentiate emotional experiences from one another.
	
	Computational work on video-induced emotions has traditionally favored categorical representation schemes \cite{Wang2015}, as does the field of automated emotion recognition in a broader sense \cite{Dmello2015}. However, psychological research has substantially criticized the theories underpinning prominent categorical representations (e.g., \cite{Ortony1990, Barrett2006}). Moreover, encoding emotions with a limited amount of categories may fail to capture differences at a level that is desired by applications, making dimensional schemes an attractive alternative for affective computing \cite{Gunes2011}. Perhaps the most prominent scheme for psychological investigations of media-induced emotions is the dimensional PAD framework \cite{Mehrabian1996}. It describes affective experiences in terms of the dimensions \textit{pleasure (P)} (is an experience pleasant or discomforting?), \textit{arousal (A)} (does it involve a high or low degree of bodily excitement?), and \textit{dominance (D)} (does it involve the experience of high or low control over the situation?). PAD has been used to code several prominent datasets for VACA research (e.g. \textit{DEAP} \cite{Koelstra2012} and \textit{MAHNOB-HCI} \cite{Soleymani2012-DB}), as well as stimuli that are widely used for emotion elicitation in psychology (e.g. the IAPS image corpus \cite{Lang1997}). As such, PAD may form a particularly good basis for incorporating insights from psychological research into VACA technology. While VACA practitioners sometimes discard the more comprehensive PAD scheme in favor of the simpler Pleasure-Arousal (PA) scheme \cite{Russel1980}, there are sound reasons for including dominance when modeling emotional responses, e.g., because it relates to central emotional appraisals \cite{Broekens2012}.
	
	Overall, there exists no consensus in VACA research about a particular way to represent emotional responses, resulting in substantial variation in approaches. While not a problem in itself, this makes it challenging to compare the psychological and technological implications of different computational research projects. Because of its relatively widespread use in VACA and the high degree of compatibility with psychological research, we adopt the PAD dimensional scheme for our data collection and modeling.
	
	\vspace{-1mm}
	\section{The Mementos Dataset} 
	\vspace{-1mm}
	In this section, we describe the creation and relevant elements of a crowd-sourced dataset that we have collected for our experiments. The Mementos Dataset contains detailed information about viewers' responses to a series of music videos, including self-reports about their emotional responses and free-text descriptions of the memories that were triggered while watching them. 
	
	\vspace{-1mm}
	\subsection{Data Collection Procedure}
	\vspace{-1mm}
	
	We recruited $300$ crowd-workers via \textit{Amazon Mechanical Turk}, each receiving $6$ \textit{USD} for their participation. After providing their informed consent about participation and data usage, subjects filled in a survey with additional information about themselves and their situation (\textit{Additional Context Measures}). Then each participant viewed a random selection of $7$ stimuli from our pool of $42$ music videos. After each video, we requested ratings for the emotions it had induced (\textit{Induced Emotion}), followed by a questionnaire asking whether watching the video had triggered any personal memories. If this was the case, viewers were required to describe each of these memories with a short text (\textit{Memory Descriptions}) and rate the feelings that they associate with these recollections (\textit{Memory-Associated Affect}). This procedure resulted in a total of $2098$ unique responses from the participants ($49$ to $50$ for each video clip). Out of these, $978$ responses from $260$ unique participants involved the recollection of at least one personal memory. For the modeling activities reported in this article, we focus only on the subset of data that triggered personal memories. See \textit{Table \ref{tab:dataCollection_overview_recollectionSubset}} for summary statistics. 
	
	\vspace{-1mm}
	\subsection{Video Stimuli}
	\vspace{-1mm}
	The dataset contains a selection of $42$ music video segments from among a set of $150$  that the researchers creating the DEAP corpus have previously evaluated for their induced affect \cite{Koelstra2012}. We opted for these videos because of two reasons. The first one is the strong capacity of musical stimuli to trigger personal memories \cite{Janata2007}. The second one is that the creators of the DEAP corpus collected PAD-ratings from multiple viewers for each evaluated video. We used these stimulus-wise ratings to balance the sample of videos that we selected for our study to produce both low and high variation responses across all three affective dimensions of the PAD-space. For example, for each video where viewers in the DEAP validation study displayed a high agreement in the pleasure dimension, we selected another one where agreement on pleasure was low). 
	
	\vspace{-1mm}
	\subsection{Self-Report Measures}
	\vspace{-1mm}
	\paragraph{Induced Emotions:} We took self-reports for emotions that participants experienced while watching the video in the form of pleasure-, arousal- and dominance-ratings on a continuous scale in the interval of $[-1, +1]$. Participants provided them using the \textit{AffectButton} instrument -- a 2d-widget displaying an iconic facial expression that changes in response to users' mouse or touch interactions. Participants can select the facial expression that best fits their affect (see \cite{Broekens2013} for a detailed description and validation). 
	
	\begin{table}
	\centering
	\vspace{1em}
	\caption{Overview of Self-Reported Data for Viewers' whose Responses to Videos involved Personal Memories}
	\vspace{-1em}
	\label{tab:dataCollection_overview_recollectionSubset}
	\small
	\begin{threeparttable}
			\begin{tabular}{llccc}
			\toprule\toprule
			Variable                                             & Measurement         &  $M$ $(SD)$  &  & $Min$/$Max$ \\ \midrule
			\textbf{Induced Emotion} & \textbf{Pleasure}   & 0.29 (0.53)  &  &  -1.00/1.00  \\
			$N=978^*$                                             & \textbf{Arousal}    & -0.02 (0.80)  &  &  -1.00/1.00  \\
			                                                      & \textbf{Dominance}  & 0.25 (0.58)  &  &  -1.00/1.00  \\
			                                                      &                     &              &  &             \\
			\textbf{Mem.-asso. Affect}     & \textbf{Pleasure}   &  0.33 (0.53)  &  &  -1.00/1.00  \\
			$N=978^*$                                            & \textbf{Arousal}    & 0.05 (0.79)  &  &  -1.00/1.00  \\
			                                                      & \textbf{Dominance}  & 0.30 (0.58)  &  &  -1.00/1.00  \\
			                                                      &                     &              &  &             \\
			\textbf{Memory Descr.}   & \textbf{Word No.}   & 25.07 (15.45)  &  &  3/103  \\
			$N=978^*$                                            &                     &              &  &             \\
			                                                      &                     &              &  &             \\
			\textbf{Personality}                                  & \textbf{Honesty}    & 2.68 (0.75)  &  &  0.50/4.00  \\
			$N=260^+$                                             & \textbf{Emotional.} & 1.96 (0.78)  &  &  0.00/3.75  \\
			                                                      & \textbf{Extravers.} & 2.54 (0.77)  &  &  0.00/4.00  \\
			                                                      & \textbf{Agreeabl.}  & 2.04 (0.67)  &  &  0.25/4.00  \\
			                                                      & \textbf{Conscien.}  & 2.62 (0.71)  &  &  0.50/4.00  \\
			                                                      & \textbf{Openness}   & 2.81 (0.66)  &  &  0.00/4.00  \\
			                                                      &                     &              &  &             \\
			\textbf{Mood}            & \textbf{Pleasure}   & 0.43 (0.40)  &  & -0.66/1.00  \\
			$N = 260^+$                                           & \textbf{Arousal}    & -0.10 (0.78) &  & -1.00/1.00  \\
			                                                      & \textbf{Dominance}  & 0.38 (0.48)  &  & -1.00/1.00  \\
			                                                      &                     &              &  &             \\
			\textbf{Demographics}                                 & \textbf{Age}        & 33.40 (6.06) &  &    25/46    \\
			$N=260^+$                                             &                     &              &  &             \\
			                                                      &                     &     $N$      &  &     $N$     \\ \cline{3-3}\cline{5-5}
			                                                      & \textbf{Gender}     &  139 female  &  &  121 male   \\
			                                                      & \textbf{National.}  &   218 USA    &  &  42 Other   \\ \bottomrule\bottomrule
			                                                      &                     &              &  &
		\end{tabular}
		\begin{tablenotes}
		\vspace{-1em}
		\smaller
		* Response-specific: measured once per response to a video \newline 
		+ Viewers-specific:  measured once per viewer 
		\end{tablenotes}
	\end{threeparttable}
	\vspace{-1em}
\end{table}

	\paragraph{Memory-Associated Affect:} We asked participants to rate the affective associations with each personal memory that a video has triggered using the AffectButton instrument. Participants could report and rate as many memories per video as they had experienced. However, only $51$ out of $978$ responses from viewers involved recollections of $2$ or more memories. For these instances, we retained only the single memory for our modeling activities with the most intense associated affect in terms of PAD-ratings.
	
	\paragraph{Memory Descriptions:} Participants were requested to describe personal memories triggered in them with a short free-text description. We requested a response in English with a minimum length of three words. After filtering for multi-recollection responses (see the previous paragraph), we retained a $978$ memory description. 
	
	\paragraph{Additional Context Measures:} Participating viewers filled in a survey that provided us with additional person- and situation-specific information. This demographic information about viewers', i.e. their \textit{age}, \textit{gender} and \textit{nationality}, as well as their personality traits in terms of the 6-factor \textit{HEXACO}-model (Honesty-Humility, Emotionality, eXtraversion, Agreeableness, Conscientiousness, and Openness to experience) collected with a brief questionnaire \cite{DeVries2013}. Finally, viewers' reported their mood at the time of participation (PAD ratings with the AffectButton).
	
	\section{Influence of Personal Memories on Video-Induced Emotions}
	\vspace{-1mm}
	In a previous study based on the Mementos dataset \cite{Dudzik2020p1}, we have demonstrated that 
	\begin{enumerate*}
		\item videos create more intensive and positive emotional responses when triggering personal memories in a viewer, and that
		\item the occurrence of and affect associated with memories explains more variability in induced emotions than theoretically relevant viewer-specific measures, e.g., personality traits  
	\end{enumerate*}.
	
\begin{figure}
		\includegraphics[width=\linewidth, trim={24em 21em 24em 20.5em},clip]{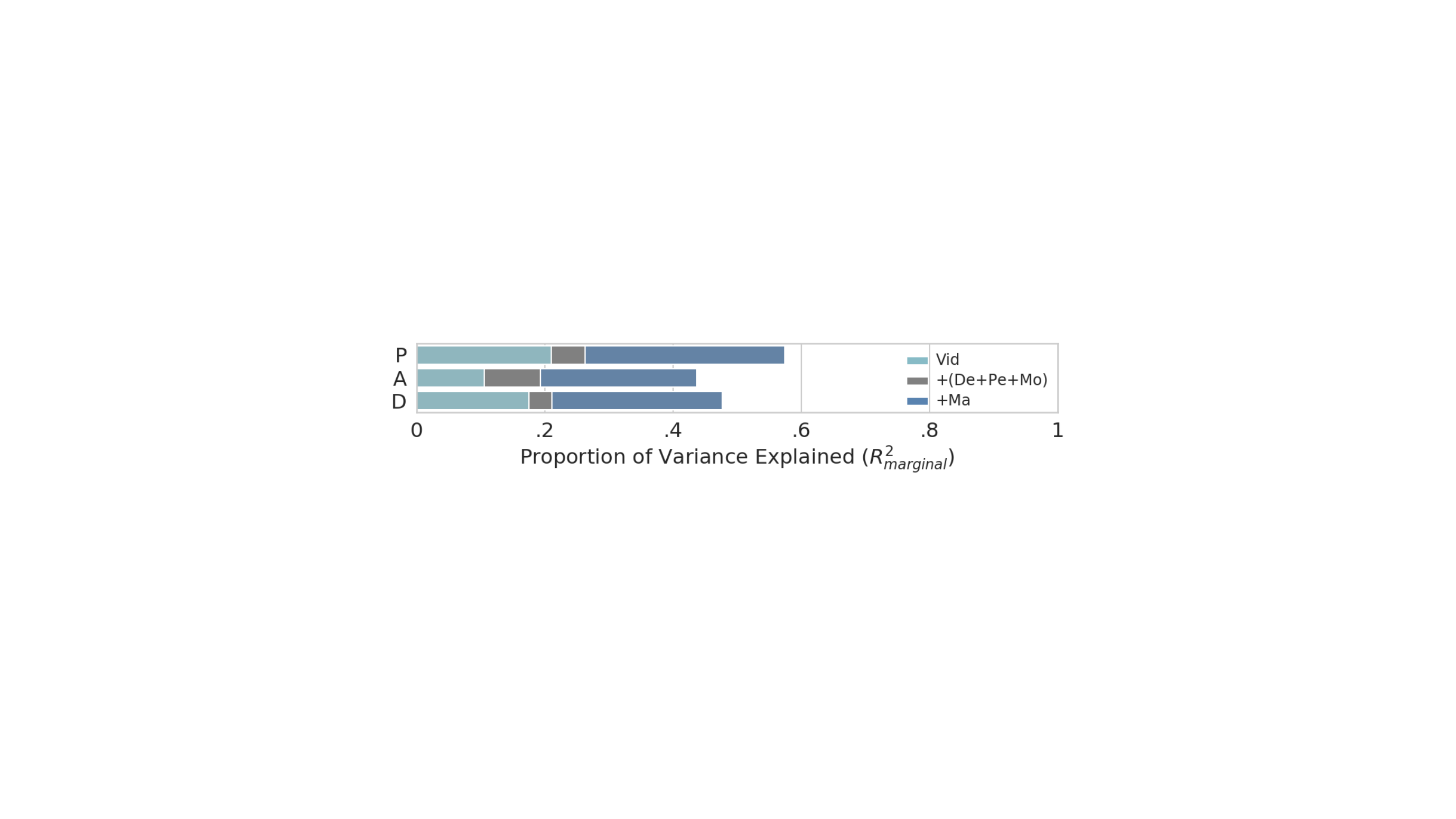}
		\vspace{-1.5em}
		\caption{Comparison of the total variance explained by different context variables for dimensions of induced emotion. The video-specific average used in context-free VACA \textit{(Vid)}. \textit{Memory-associated affect (+Ma)} accounts for a greater share of additional uniquely explained variance than \textit{demographics}, \textit{personality} and \textit{mood} combined \textit{(De+Pe+Mo)}.}
		\label{fig:sec4_ContextVariableComparison}
		\vspace{-1.5em}
	\end{figure}

	To further illustrate the relevance of personal memories as context for predicting individual viewers' experiences, we extend these earlier investigations here with a focused statistical analysis of only those responses that involve memories. Concretely, we quantify the variance explained by memory-associated affect in comparison to that of viewer-specific context variables captured in the dataset (see Demographics, Personality, and Mood in \ref{tab:dataCollection_overview_recollectionSubset}). For this purpose, we conduct a regression analysis with nested linear mixed-effects models targeting the emotions induced in viewers (one model per affective dimension: Pleasure (P), Arousal (A), and Dominance (D)). \textit{Figure \ref{fig:sec4_ContextVariableComparison}} shows the differences in explained variance by these models for \begin{enumerate*}
		\item a baseline model ($Vid$), predicting only the video-specific average (i.e. the \textit{Expected Emotion} in context-free VACA), 
		\item the combined contribution of \textit{Demographics}, \textit{Personality}, as well as \textit{Mood} as additional predictors in the model ($+(De+Pe+Mo)$, and finally 
		\item the additional effect of \textit{memory-associated affect} ($+Ma$). 
	\end{enumerate*}
	
	We see that the combined viewer-specific measures account for a significant share of additional variance over the baseline (P: ${\Delta}R^2_{m} =.052$, $F(13, 213.55)=4.05$, $p<.001$; A: ${\Delta}R^2_{m} = .088$, $F(13, 219.21)=5.75$, $p<.001$; D: ${\Delta}R^2_{m} = .036$, $F(13, 213.68)=2.85$, $p<.001$). However, information about memory-associated affect explains a large share of unique variation on top of that (P: ${\Delta}R^2_{m} = .364$, $F(16, 268.76)=48.1$, $p<.001$; A: (${\Delta}R^2_{m} = .332$, $F(16, 278.06)=31.86$, $p<.001$); and D: (${\Delta}R^2_{m} = .301$, $F(16, 263.22)=263.22$, $p<.001$). In fact, the average amount of additional variance explained by memory-associated affect for affective dimensions is about \emph{six times} higher. Together, these findings provide extremely compelling evidence for the exploitation of video-triggered memories as a form of situation-specific context for predictions in VACA.    
	
	\vspace{-1mm}
	\section{Predictive Modeling}
	\vspace{-1mm}
	In this section, we outline the framework that we use as a proof of concept for integrating descriptions of personal memories triggered by a video as contextual information. We have made essential design choices, in how we collected our data to stay true to recognized affective theory and its relation to the task at hand. As a consequence, no state-of-the-art baseline exists from which we can compare our approach to others. We do \emph{not} intend to provide significant extensions to the technological state-of-the-art. Instead, we provide a state-of-the-art baseline to investigate the contribution of memory descriptions such that we can better expose the nature of this novel task approach. Concretely, we describe the machine learning models and multimodal fusion strategies that we deploy for this task.

	\vspace{-1mm}
	\subsection{Overview}
\begin{figure}
		\includegraphics[width =\linewidth, trim={4em 62em 4em 3.5em},clip]{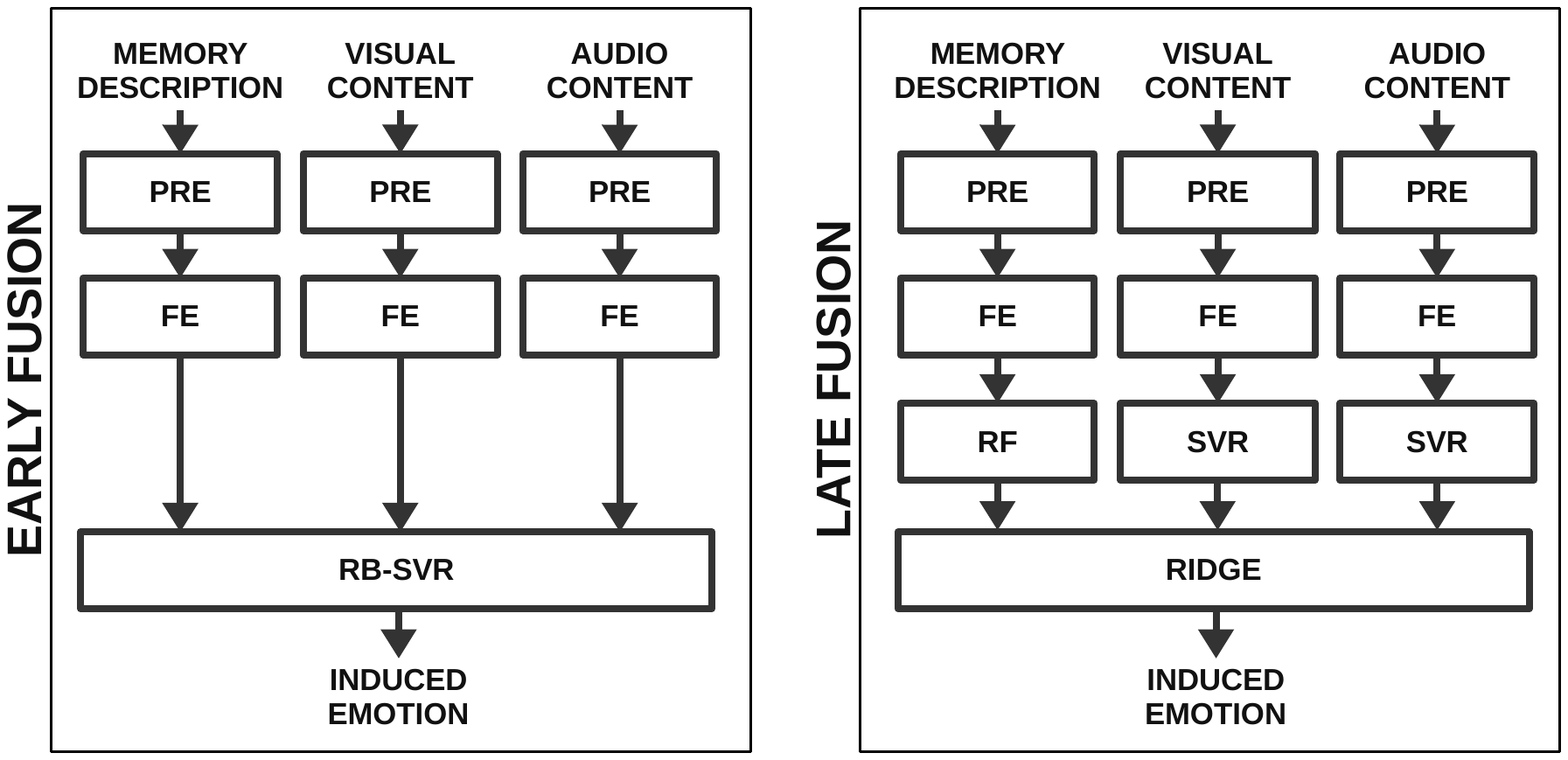}
		\vspace{-2.5em}
		\caption{Overview of our approaches for predictive modeling and multimodal fusion. \textit{PRE:} Preprocessing; \textit{FE:} Feature Extraction;  \textit{SVR:} Support Vector Regression; \textit{RF:} Random Forest Regression; \textit{RIDGE:} L2-regularized linear model}
		\label{fig:sec5_fusionApproaches}
		\vspace{-1.5em}
	\end{figure}

	\vspace{-1mm}
	In line with previous work on VACA (\cite{Wang2015}), we address modeling viewers' emotional responses as a regression problem. An important aspect of context-sensitive VACA technology is the integration of information from different sources into a single multimodal prediction model, i.e., video content and context features. Previous VACA work has repeatedly relied on early fusion to combine the information provided by different sources. Support Vector Machines are a popular type of machine learning algorithm for this purpose (e.g., \cite{Scott2016, McDuff2017}, especially when predictions take place in a regression setting (see \cite{Wang2015}). We also consider the state of the art approaches for emotion prediction from text since we intend to exploit free-text descriptions of memories as context in predictions. Interestingly, state-of-the-art results for predicting the author's affective states from short texts (i.e., tweets) have been achieved by decision-level fusion. For instance, Duppada et al. used random forest regressors with a regularized linear model as a meta regressor different text feature-sets \cite{Duppada2018}. Motivated by this, we explore both early and late fusion approaches in our experiments. \textit{Figure} \ref{fig:sec5_fusionApproaches} provides a graphical overview of the two machine learning pipelines that we deploy: an \textit{Early Fusion-Approach}, consisting feature-level fusion of the different information sources, and a \textit{Late Fusion-Approach}, using a meta-regressor to combine the predictions from modality-specific models via stacking. For all machine learning algorithms, we used the implementation provided by the \textit{Scikit-Learn} python library \cite{Pedregosa2011} in Version $0.22.2$.    
	
	\paragraph{\textbf{Early Fusion-Approach:}} This pipeline consists of a series of pre-processing and feature extraction steps that are specific for memory, visual, and audio sources. The resulting feature vector is then concatenated, and fed to a Support Vector Regressor (SVR) with Radial Basis-Function (RBF) kernel for predictions.
	
	\paragraph{\textbf{Late Fusion-Approach:}} Here, we also perform the same modality-specific pre-processing and feature extraction operations. We then rely on two separate SVRs with RBF kernel for predictions based on the audio or visual features, while, in line with \cite{Duppada2018}, we use a Random Forest Regressor for predictions based on memory descriptions. We combine the output of the modality-specific models with an L2-regularized linear regression model (Ridge regression).
	
	\vspace{-1mm}
	\subsection{Stimulus Video Processing}
	\vspace{-1mm}
	To represent the audio content of the music videos, we used the software \textit{openSMILE} in the configuration \textit{“emobase2010"} for feature extraction. It derives low-level descriptors from raw audio signals in a windowed fashion and aggregates them statistically, resulting in a $1582$-dimensional feature vector (see \cite{Schuller2010} for a detailed description). This feature set is widely used for audio representation in VACA research as a baseline approach in benchmarking challenges \cite{Delleandrea2018}.
	
	For visual representation of the stimulus video, one frame is extracted every second. For each frame, we then extract three types of visual features: 
	\vspace{-1mm}
	\paragraph{\textbf{Theory-inspired Descriptors:}} Work on affective visual content analysis has developed features that were specifically engineered to capture the affective properties of images. These descriptors often are inspired by findings from psychological research or art-theoretic concepts. We use the set of descriptors developed by Machajdik \& Hanburry \cite{Machajdik2010}, as well as those of Bhattacharya et al. \cite{Bhattacharya2013} to characterize each of the extracted video frames (resulting in a $271$-dimensional feature vector). This combination has been used in context-sensitive VACA work before \cite{Scott2016}. 
	
	\vspace{-1mm}
	\paragraph{\textbf{Deep Visual Descriptors:}} Deep neural networks form an essential part of modern approaches to visual content analysis and computer vision. Instead of engineering descriptors, these models learn effective and reusable representations for prediction tasks directly from visual training data. We use the activation of the FC1-layer of a pre-trained VGG16 network \cite{Simonyan2015} from the Keras framework for python \cite{Chollet2015} for this purpose (resulting in a feature vector with 4096 dimensions). This representation has been used for visual representation in VACA research as a baseline in benchmarking challenges \cite{Delleandrea2018}.
	
	\vspace{-1mm}
	\paragraph{\textbf{Visual Sentiment Descriptors:}} Classifiers that identify the presence of \textit{Adjective-Noun Pairs (ANPs)} in images have been successfully used as high-level descriptors in VACA approaches (e.g., \cite{Scott2016, McDuff2017}). ANPs consist of labels denoting objects or persons identified in an image, coupled with an affective attribute (e.g., "beautiful house"). We use the class-probabilities assigned by the \textit{DeepSentiBank} Network \cite{Chen2014} for any of the ANPs in its ontology as descriptors for the content of video frames ($4342$-dimensional feature vector). 
	
	We concatenate these different feature sets into a combined 8709-dimensional vector to represent the visual content of individual extracted video frames. We compute these for each of the frames extracted from a video, and then take the dimension-wise average across them to produce a single $8709$-dimensional representation of the entire stimulus video's visual content.
	
	\vspace{-1mm}
	\subsection{Memory Description Processing}
	\vspace{-1mm}
	We preprocess memory descriptions by replacing references to specific years or decades (e.g. "1990", or "the 90s") with generic terms (e.g. "that year" or "that decade"). Additionally, we replace any numbers with $0$ and expand all contractions present in participants' descriptions (e.g. "can't" is transformed into "cannot"). To model the affective impact of personal memories we rely on features that have proven successful in state-of-the-art work modeling emotional states from social media text in a regression setting \cite{Mohammad2018}: \begin{enumerate*} \item \textit{Lexical Features} and \item \textit{Word Embeddings} \end{enumerate*}.
	
	\vspace{-1mm}
	\paragraph{\textbf{Lexical Features:}} We generate these features by parsing memory descriptions into word-level tokens, for which we then retrieve associated affective ratings from a wide variety of affective dictionaries. To account for differences between words used in memory descriptions and the form in which they are typically indexed in lexica, we apply lemmatization before the lookup to remove inflections. The combination of the dictionaries that we initially selected for feature extraction \cite{Hu2004, Baccianella2010, Mohammad2010, Thelwall2010, Tausczik2010, Nielsen2011, Mohammad2013, Choi2014, Mohammad2015, BravoMarquez2016, Mohammad2017} was demonstrated to contribute to state-of-the-art performance for affective text regression tasks \cite{Duppada2018}. We extended this list by a recent addition that provides word-level ratings for Pleasure, Arousal, and Dominance \cite{Mohammad2018}. We aggregate the associated word-level ratings from each dictionary by averaging the extracted features for each word token in a memory description. Additionally, we include the sentiment scores provided by the \textit{VADER} model \cite{Hutto2014} when applied to an entire memory description. It combines an empirically collected sentiment lexicon with a set of rule-based processing steps to score the affective valence of text. Together this results in a $130$-dimensional vector of lexical features for each description. 
	
	\vspace{-1mm}
	\paragraph{\textbf{Word Embeddings:}} We leveraged two pre-trained word embedding-models to represent each word in the memory description texts as a real-valued feature vector:  \textit{(1)} \textit{Word2Vec}-model pre-trained on the \textit{Google News dataset}, resulting in a $300$-dimensional feature vector when applied to a word, and \textit{(2)} a \textit{GloVE}-model \cite{Pennington2014} pre-trained on the \textit{Wikipedia 2014 and the Gigaword 5 corpora}. It encodes individual words as a $200$-dimensional feature vector. For both implementations we rely on the \textit{Gensim}-library for python \cite{Rehurek2010}. To generate a representation for the entire memory description from these word-level embeddings, vectors of both types are concatenated and averaged for the entire description, resulting in a single $500$-dimensional feature vector for each memory.

	\vspace{-1mm}
	\section{Empirical Investigations}
	\vspace{-1mm}
	We conduct two experiments in an ablation setting of our early and late fusion approaches when predicting the video-induced emotions of viewers in the Mementos Dataset. In \textit{Experiment 1}, we assess the feasibility of predicting viewers' emotional responses to a video from text describing the memories that it has triggered in them. \textit{Experiment 2} quantifies the relative contribution of memory descriptions when used for predicting emotional responses alongside the audiovisual content of videos. 
	
	\vspace{-1mm}
	\subsection{Experimental Setup and Evaluation}
	\vspace{-1mm}
	We use a nested 5-Fold-Leave-Persons-Out Cross-Validation for training and evaluation of our early and late fusion approaches. In this procedure, folds are created such that no data from the same individual is spread across training and validation. In the outer loop of the nested cross-validation, we split the entire dataset into $5$ folds, from which we hold out a single fold for testing the final performance of selected models. In the inner loop, the remaining $4$ folds are used for selecting hyperparameters for the machine learning models with a grid search. 
	
	\vspace{-1mm}
	\subsection{Results and Analysis}
	\vspace{-1mm}
	
	\subsubsection{Experiment 1 -- Using Memory Descriptions for Prediction}
\begin{figure}
		\includegraphics[width =\linewidth, trim={21.7em 14em 21.7em 14em},clip]{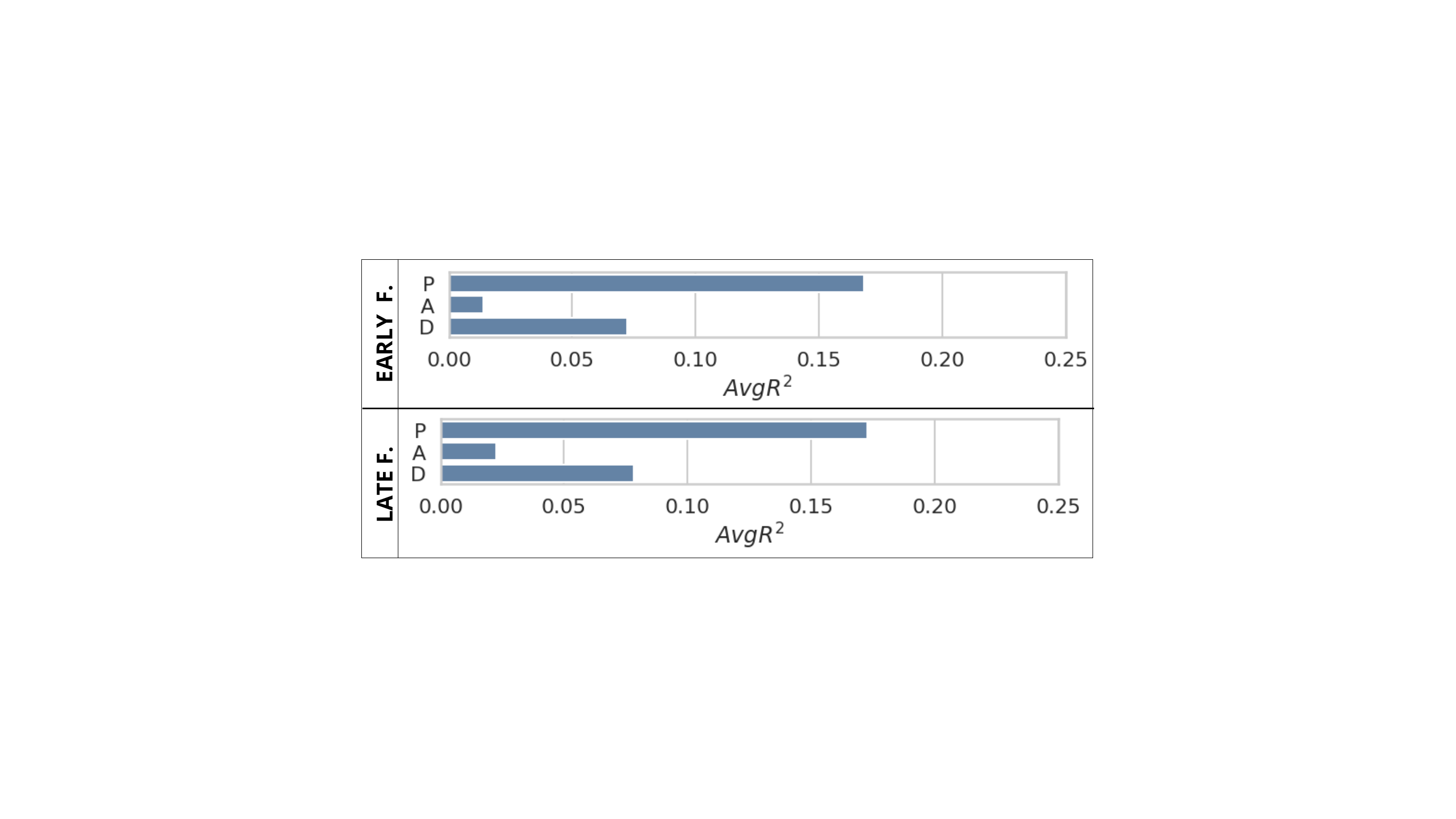}
		\vspace{-2em}
		\caption{Results for Experiment 1 --- Average Test-Performance for Early and Late Fusion Approaches for predicting viewers' induced \textit{Pleasure (P)}, \textit{Arousal (A)}, and \textit{Dominance (D)} when using only Memory Descriptions.}
		\label{fig:sec6_Exp1_modelComparisons}
		\vspace{-1em}
	\end{figure}

	We show the average performance of our early and late fusion approaches when only provided with viewers' memory descriptions as input in \textit{Figure \ref{fig:sec6_Exp1_modelComparisons}}. These findings indicate that text descriptions explain a significant portion of the variance in induced pleasure and, to a lesser degree, dominance (i.e., their $AvgR^2 > 0$). However, the performance of our models is much lower for arousal.  
	
	To explore whether this decrease is a result of our modeling choices, we investigate how well humans can infer affective information from the memory descriptions in our dataset. For this purpose, two raters manually annotated a random selection of $150$ descriptions with two kinds of affective evaluations for pleasure, arousal, and dominance: \begin{enumerate*}
		\item -- \textit{Perceived Conveyed Affect (PCA)} of the text, and
		\item the \textit{Inferred Affective Experience (IAX)} of the author.  
	\end{enumerate*}
	For PCA ratings, the annotators answer the question \textit{"What feelings does this text express?"}. We instruct them only to consider emotions or feelings that are explicitly described by the authors, e.g., by using emotion words like "love" or "hate".  Performance on this task will provide us with insights about how explicit authors describe their emotions in the text. In the case of IAX ratings, annotators answer the question \textit{"How do you think the person describing this memory feels about it? Put yourself into their situation".} The motivation for the different task formulation is to encourage annotators to draw on their cultural background and experience to infer implicit emotional meaning from the descriptions. Such inferences are a vital component of human emotion perception \cite{Barrett2011}, and performance on this task shows the degree to which the texts facilitate them. Raters provide PAD annotations with the widely adopted and validated \textit{Self-Assessment Manikin} instrument \cite{Bradley1994}. Based on this information, we assess the \textit{correspondence} of the two annotators' ratings with viewers' self-reported ratings for memory-associated affect (MA), and the \textit{reliability} with which they were able to do so. For this, we calculate Pearson correlations between the average PCA/IAX ratings of both annotators and viewers' MA ratings to measure correspondence. Similarly, we calculated the correlations between the PCA/IAX ratings between both raters as a measure of reliability and agreement. Results for both are listed in \textit{Table \ref{tab:analysis_results_corAnnotations}}). Our findings show that annotators' ratings for pleasure in the PCA and IAX tasks both highly agree with viewers' experienced emotions, as well as each other. This pattern is still present -- albeit less strongly pronounced -- for dominance. However, for arousal, annotators' judgments correspond much less with viewers' MA ratings. Moreover,  annotators also tend to disagree much more with each other. On average, both correspondence and reliability of judgments are higher in the IAX task than for the PCA task. This result confirms our hypothesis that simulating the authors' state of mind helps human annotators. However, levels of performance do still not reach those displayed for pleasure or dominance. This finding points towards an inherently greater difficulty for recognizing arousal from text, rather than to a particular weakness in our modeling approach. A likely explanation is that descriptions contain few explicit or implicit expressions of an author's arousal, making it challenging for both humans and automatic text analysis to perform well.

	\subsubsection{Experiment 2: Combining Direct VACA and Memory Context}
	\begin{figure}
		\includegraphics[width =\linewidth, trim={21.7em 6.8em 21.7em 6.8em},clip]{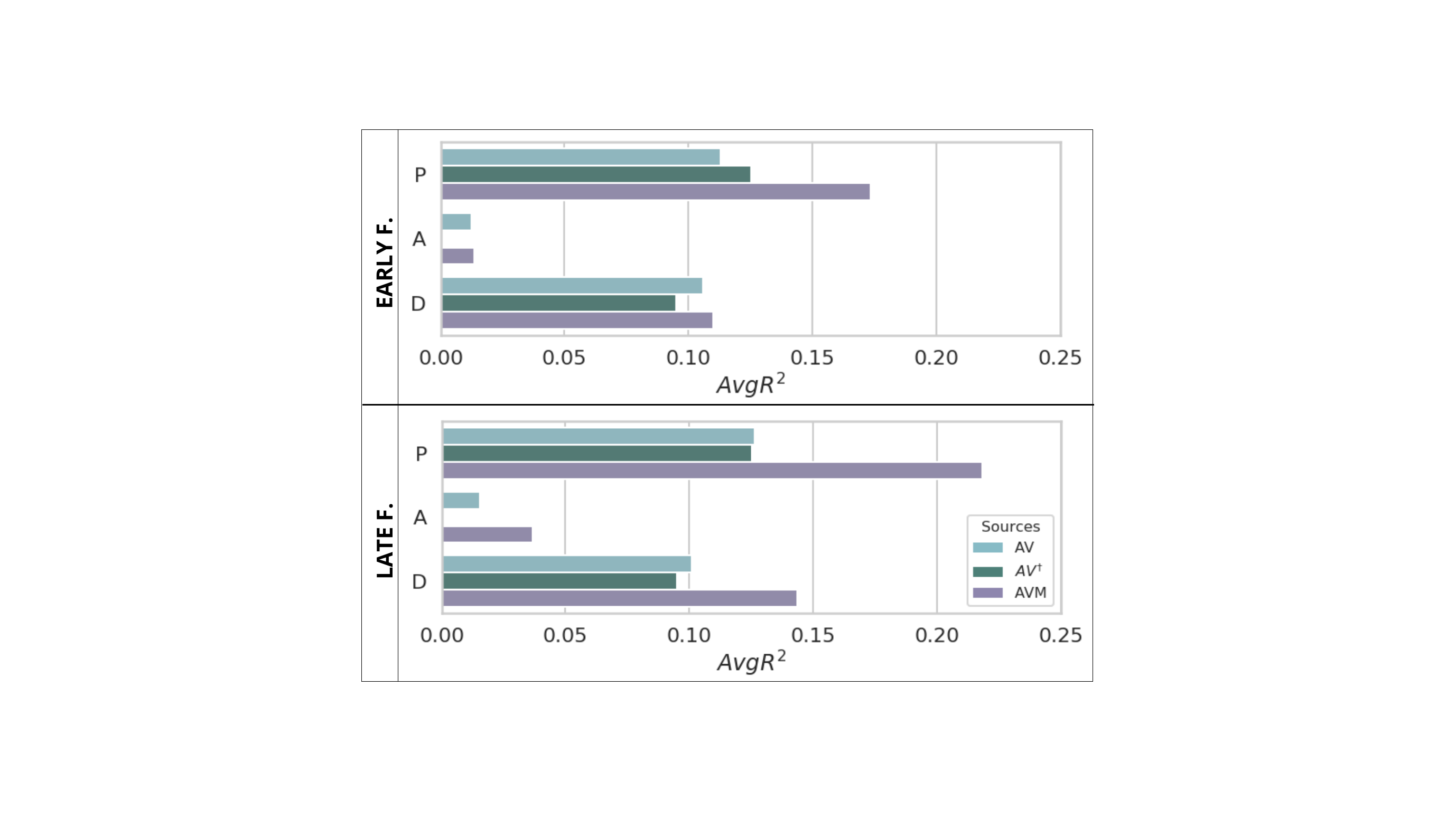}
		\vspace{-2em}
		\caption{Results for Experiment 2 --- Average Test-Performance for Early and Late Fusion Approaches when predicting induced \textit{Pleasure (P)}, \textit{Arousal (A)}, and \textit{Dominance (D) using different sources. $AV$: audio-visual tracks; $AV^{\dag}$: Model approximating optimal results for Context-Free VACA using the video-specific mean of the ground truth; $AVM$: audio-visual tracks combined with memories.}}
		\label{fig:sec6_Exp2_modelComparisons}
		\vspace{-1em}
	\end{figure}
	
	\begin{table}
	\centering
	\vspace{1em}
	\caption{Human Annotators' Performance for Affective Interpretation of Memory Descriptions (Pearson Correlations)}
	\vspace{-1em}
	\label{tab:analysis_results_corAnnotations}
	\small
	\begin{threeparttable}
        \begin{tabular}{ccccccccc}
            \toprule
            \toprule
            & \multicolumn{3}{c}{Correspondence} &  &     & \multicolumn{3}{c}{Reliability} \\
            & P             & A             & D            &  &     & P            & A           & D           \\
            PCA & .58***        & .15           & .39***       &  & PCA & .78***       & .18*        & .54***      \\
            IAX & .55***        & .26***        & .44***       &  & IAX & .86***       & .31**       & .58***    \\
            \bottomrule
            \bottomrule
        \end{tabular}
		\begin{tablenotes}
			* $p <= .05$; 
			** $p <= .01$;
			*** $p <= .001$
		\end{tablenotes}
	\end{threeparttable}
	\vspace{-1em}
\end{table}
	
	\textit{Figure \ref{fig:sec6_Exp2_modelComparisons}} shows the performance displayed by our approaches when having access to either 
	\begin{enumerate*}
		\item only videos' audiovisual data ($AV$), or
		\item a combination of both audiovisual data and memory descriptions ($AVM$). 
	\end{enumerate*}     
	In addition, we also list the performance of a model that predicts the video-wise mean for affective dimensions of induced emotions from a sample ($AV_{\dag}$). This measure indicates the best-case performance that a context-free VACA model can provide for the current dataset if it always makes the correct prediction of the "Expected Emotion" for a video.
	
	Our analysis shows that memory descriptions can provide substantial additional information about viewers' affective responses, independent of fusion strategy. Overall, improvements when giving models access to memory descriptions are most pronounced for predictions of pleasure (Early Fusion: ${\Delta}Avg{R^2} = +.059$; Late Fusion: ${\Delta}Avg{R^2} = +.092$;). This evidence is consistent with the findings of experiment 1, which demonstrates the greater ability of our models for predicting pleasure. Similarly, the performance for arousal remains poor across the board, even when using audiovisual features only. Despite the comparably simple modeling approach that we have deployed, the performance of our models using only audiovisual features ($AV$) approximates that offered by an ideal context-free VACA model ($AV\dag$) (absolute differences in performance averaged across affective dimensions -- Early Fusion: ${\Delta}Avg{R^2} = .003$; Late Fusion: ${\Delta}Avg{R^2} = .006$). Finally, we observe that the late fusion approach displays a two times greater increase in performance when provided with memory descriptions than the early fusion approach (difference in performance gains averaged across affective dimensions -- Early: ${\Delta}Avg{R^2} = +.021$; Late: ${\Delta}Avg{R^2} = +.054$). This result highlights that despite the efficacy of early fusion in classic VACA, research should not rule out late fusion for context-sensitive approaches.
	
	\balance
	
	\vspace{-1mm}
	\section{Discussion}
	\vspace{-1mm}
	Our empirical investigations demonstrate that it is both feasible to use viewers' self-reported memory descriptions for predicting emotional experience and that doing so provides a valuable source of context for personalized predictions in video affective content analysis. Particularly for pleasure, the automatic analysis of memory descriptions explains variation that is substantially above and beyond that of a video's audio and visual content. Surprisingly, none of our models provided substantial insights into viewers' arousal. Our findings that even humans struggle to reliably and accurately infer arousal from memories offer a possible explanation. Nevertheless, the weak performance for arousal when using stimulus features is surprising, since previous research achieved performance of arousal predictions comparable to pleasure or dominance (e.g., \cite{Koelstra2012, Yazdani2013a, Baveye2015}). A reason for this might be that we did not filter our stimuli based on the variability of responses they elicited before modeling. Especially for arousal, our viewers reported widely different levels of arousal to the same video. This fluctuation might make it harder for algorithms to learn directly from a video's audiovisual content. Similarly, the stimulus-specific mean as "Expected Emotion" might not be a good approximation for these cases. This last finding underlines the limited capacity of purely context-free predictions to reflect viewers' individual experiences of video stimuli accurately.
	
	Overall, the studies demonstrate that descriptions of video-triggered personal memories are a viable and useful resource for making personalized predictions. However, there are some limitations to our findings. First, the memory descriptions that we analyzed may differ from those that viewers would create in-the-wild, such as on social media platforms, or during a conversation. Second, we have collected them from paid crowd-workers, who provided them with the full knowledge that their identity remains protected. Under these circumstances, participants may have been willing to provide more detailed and candid accounts of their memories than they otherwise would have. Future work could expand on our findings by compiling corpora that capture how people describe media-evoked recollections under more natural conditions in the wild. An effective way to achieve this might be to elicit recollections over more extended periods with specifically developed applications, e.g., via social media \cite{Cosley2009}, or through conversations with interactive intelligent agents \cite{Peeters2016}. Moreover, research could explore technological efforts to identify and extract memory descriptions from generally available data -- e.g., social media posts. -- automatically. 
	
	However, relying on the availability of explicit descriptions can only be a first step for improving predictions in real-world settings. For many use-cases of VACA, such descriptions may not be available at prediction time, as viewers can only describe the memories triggered in them by a video after they have already been exposed to it, but not before. To anticipate the effect of memories on viewers beforehand -- e.g., to personalize recommendations of unseen content --, models will have to estimate when memories are triggered and what their content will be. While these are undoubtedly challenging tasks, progress towards achieving them in automated systems seems feasible. A first step could be to explore how well the use of already obtained text-based memory descriptions generalizes to new, but related stimuli, e.g. music videos from the same artist. Additionally, Dudzik et al. \cite{Dudzik2018} argue that existing research from ubiquitous computing and cognitive modeling offers numerous starting points for modeling memory processes in adaptive media technology. For example, they propose that modeling a user's attentional focus from sensor data might be a way to identify situations and stimuli that are likely to trigger memories. Given the substantial role that personal memories play for induced emotions, such research topics should be actively pursued in context-sensitive VACA. Naturally, this requires access to rich corpora of data that facilitate such investigations. In this spirit, we plan to publish the Mementos dataset as a resource for future work on modeling affective memory processing during video consumption.
	
	\vspace{-1mm}
	\section{Summary and Conclusion}
	\vspace{-1mm}
	Video-affective Content Analysis (VACA) has traditionally operated under the assumption that videos possess a more or less objective emotional meaning, existing across different viewers and independent of the situation under which they are experienced. However, the emotional impact of videos in the real world is highly subjective and varies greatly with the situation. Consequently, not accounting for context limits the ability of predictions to reflect emotional responses accurately. Contemporary video content is consumed by an increasingly diverse, global-spanning community and in a broad variety of circumstances. Given these developments, research on context-sensitive approaches to VACA seems vital for predictions to be of use to media applications. 
	
	Personal memories that are triggered in viewers form a highly situation-specific form of context, shaping individual emotional responses to the video. The two empirical investigations that we describe in this article show the feasibility of using free-text descriptions to account for this influence in automated predictions. Moreover, our findings demonstrate that combining this approach with an analysis of a video's audiovisual content can provide significant performance benefits. As such, when memory descriptions are available, they offer a powerful form of context for personalizing predictions in VACA. Because people tend to talk about their memories with each other, automatic approaches can feasibly extract such descriptions from communications on social media or face-to-face interactions. Nevertheless, the investigations described in this article only form a first step towards accounting for the influence of memories in automatic predictions. Future investigations should more generally explore modeling the occurrence (when?), content (what?), and influence (what does it do?) of personal memories in VACA predictions to support media applications in the real world.
	
	
	\bibliographystyle{ACM-Reference-Format}
	
	\bibliography{bibliography/P2_ARXIV.bib}

\end{document}